\documentclass[fleqn,usenatbib]{mnras}

\usepackage{newtxtext,newtxmath}

\usepackage[T1]{fontenc}

\DeclareRobustCommand{\VAN}[3]{#2}
\let\VANthebibliography\thebibliography
\def\thebibliography{\DeclareRobustCommand{\VAN}[3]{##3}\VANthebibliography}

\usepackage{graphicx}	
\usepackage{amsmath}	
\usepackage{gensymb}	

\newcommand{\beq}{\begin{equation}}
\newcommand{\eeq}{\end{equation}}

\def\dd\mathrm{d}

\def\xbr{x_{\mathrm{break}}}

\title[The eFEDS Galaxy Cluster Power Spectrum]{The X-ray Angular Power Spectrum of Extended Sources in the eROSITA Final Equatorial Depth Survey}

\author[Lau et al.]{Erwin T. Lau,$^{1,2}$\thanks{E-mail: erwin.lau@cfa.harvard.edu}
\'Akos Bogd\'an,$^{1}$
Urmila Chadayammuri,$^{1}$
Daisuke Nagai,$^{3}$
Ralph P. Kraft,$^{1}$ and
\newauthor
Nico Cappelluti$^{2}$
\\
$^{1}$Center for Astrophysics | Harvard \& Smithsonian, 60 Garden St, Cambridge, MA 02138, USA\\
$^{2}$Department of Physics, University of Miami, Coral Gables, FL 33124, USA\\
$^{3}$Department of Physics, Yale University, New Haven, CT 06520, USA \\
}

\date{Accepted: 2022 October 16; Revised: 2022 October 12; Received: 2022 April 12}

\pubyear{2022}

\begin{document}
\label{firstpage}
\pagerange{\pageref{firstpage}--\pageref{lastpage}}
\maketitle

\begin{abstract}
The {\em eROSITA} Final Equatorial Depth Survey (eFEDS), with a sky area of 140 square degrees with depth equivalent to the equatorial patch of the final {\em eROSITA} all-sky survey, represents the largest continuous non-full-sky X-ray fields to-date, making it the premier data set for measuring the angular power spectrum. In this work, we measure the X-ray angular power spectrum of galaxy clusters and groups in the eFEDS field. We show that the measured power spectrum is consistent with past observations, including the ROSAT All Sky Survey, and the {\em Chandra} COSMOS and Bootes fields.  The predictions of cluster gas halo model that is calibrated from Chandra observations is also consistent with the eFEDS power spectrum. While the eFEDS does not have large enough sky coverage to provide meaningful cosmological constraints, we predict that the X-ray power spectrum from the cycle 4 of the {\em eROSITA} All-Sky Survey (eRASS4) will provide constraints on $\Omega_M$ and $\sigma_8$ at the $10\%$ level. \end{abstract}




\section{Introduction}

While the standard $\Lambda$ cold dark matter ($\Lambda$CDM) model has been successful describing a number of observations, from the primary anisotropies of the cosmic microwave background (CMB), baryon acoustic oscillations (BAO), and standard candles such as supernova Ia, recent observations gradually reveal some unsettling and tantalizing issues. One such issue is the so-called ''$S_8$'' tension. The quantity $S_8 \equiv \sigma_8(\Omega_M/0.3)^{1/2}$ is a combination of two key cosmological parameters, $\sigma_8$ and $\Omega_M$, which represent the fluctuations and the amount of matter in the Universe, respectively. Low-redshift cosmological probes, specifically, cosmic shear galaxy clustering, and abundances of galaxy clusters  \citep{cfhtls, kids450, kids1000_lens, des_y3, des_y1_cl}, all gives consistently low values of $S_8$, compared to value inferred from the primary anisotropies of the CMB, measured by {\em Planck} \citep{planck2018} at a level of $3\sigma$. One possible solution is the suppression of power on small scales \citep{amon_efstathiou22} by strong baryonic feedback, or more exotic physics like neutrinos. 

Being the largest and most massive virialized structures in the Universe, galaxy clusters are unique in that they probe both astrophysics and cosmology at the same time. In particular, the angular power spectrum of galaxy clusters offer a promising avenue to address the $S_8$ tension by  simultaneously constraining baryonic physics and cosmological parameters. The amplitude of the angular power spectrum is especially sensitive to $S_8$ \citep{komatsu_seljak02}. Unlike conventional cluster abundance measurements, angular power spectrum does not require accurate and precise measurements of galaxy cluster masses and their selection \citep{pratt_etal19}, thus it is subject to different astrophysical systematics compared to conventional cluster abundance measurements \citep[e.g.,][]{battaglia_etal10, shaw_etal10}. As the angular power spectrum measures the clustering of galaxy clusters and groups over a wide range of masses and redshifts. It also allows us to probe cluster astrophysics that are otherwise difficult to constrain with individual cluster observations, especially for less massive and high redshift groups that cannot be resolved with current instruments. The X-ray or thermal Sunyaev–Zeldovich Effect (SZ) angular power spectrum can be measured without selection of clusters or deriving their masses by probing the X-ray or SZ clustering signals over a given observed patch of the sky. However, the clustering signal can be contaminated by foreground and background sources. 

The angular power spectra in X-ray and SZ are sensitive to cluster astrophysics, such as the physics in the outskirts of clusters \citep{walker_etal19}, in particular the non-thermal pressure and gas density clumping. These physics are difficult to constrain with individual cluster observations, as the outskirts have low signal-to-noise both in X-ray and SZ given the low gas density and pressure.  Angular power spectra, on the other hand, are sensitive to these physics as they measure the total fluctuations of gas properties imposed by these physics at scales corresponding to halo outskirts \citep{shirasaki_etal19}. 

The angular power spectrum of galaxy clusters has been measured in the microwave band via the thermal SZ Effect, in which CMB photons are scattered off the hot electron in clusters, resulting in the spectral distortion in the CMB photons. The analysis of the angular power spectrum of the tSZ maps of the {\em Planck} mission had yielded cosmological constraints that are complimentary to cluster abundance measurements \citep{planck_sz_power}. The cluster angular power spectrum is also promising in probing neutrino mass \citep[e.g.][]{bolliet_etal19}, and models of dark matter, such as sterile neutrinos  \citep{zandanel_etal15}. 

The first measurement of the X-ray angular power spectrum of clusters and groups was performed by \citet{diego_etal03}, who used the ROSAT all sky survey to place an upper limit on $\sigma_8$. Cosmological constraints were also derived by cross-correlating the ROSAT all-sky map with the {\em Planck} tSZ map \citep{hurier_etal15}, subjected to uncertain astrophysics such as non-thermal pressure \citep{shaw_etal10}. 

The recently released X-ray data from the eROSITA Final Equatorial-Depth Survey \citep[eFEDS;][]{eFEDS} presents a unique data set to provide the first X-ray power spectrum constraints on both cosmology and astrophysics. Previous X-ray power spectrum measurements with {\em Chandra} \citep{kolodzig_etal17,kolodzig_etal18} covered only small patches of the sky which lacked the range of angular scales to simultaneously probe both cosmology and cluster astrophysics. The eFEDS field has the largest sky area coverage to date, except the now 30 yr old all-sky ROSAT data \citep{snowden_etal97} which has $30$--$50$ times lower sensitivity. Before the public release of the eROSITA all sky survey, it remains the best data set to measure the X-ray power spectrum over a larger range of angular scales. Using eFEDS is also preferable to combining survey data from different X-ray instruments ({ROSAT}, XMM--{\em Newton}, {\em Chandra}), which have different systematics. 

In this paper, we provide the first measurement of the X-ray angular power spectrum of the eFEDS field. We show that the eFEDS X-ray power spectrum is consistent with a model of the X-ray emissions of clusters and groups that is calibrated with independent measurements of density profiles and gas masses of galaxy clusters and groups.
While the eFEDS data is still not adequate for providing competitive constraints in cluster astrophysics and cosmology, we show that the constraint will be improved with the all-sky {\em eROSITA} data. 

The paper is organized as follows.  In Section~\ref{sec:data}, we present and describe the eFEDS data analysis.  In Section~\ref{sec:model}, we describe our model of the X-ray angular power spectrum model. In Section~\ref{sec:results}, we present our comparison between our model and observations. We provide a summary of our results and discussion in Section~\ref{sec:summary}. 

Throughout this paper, we assume {\em Planck}18 cosmology \citep{planck2018}, unless noted otherwise. 

\section{Analysis of the eFEDS Data}
\label{sec:data}

\begin{figure*}
\centering
\includegraphics[scale=0.7]{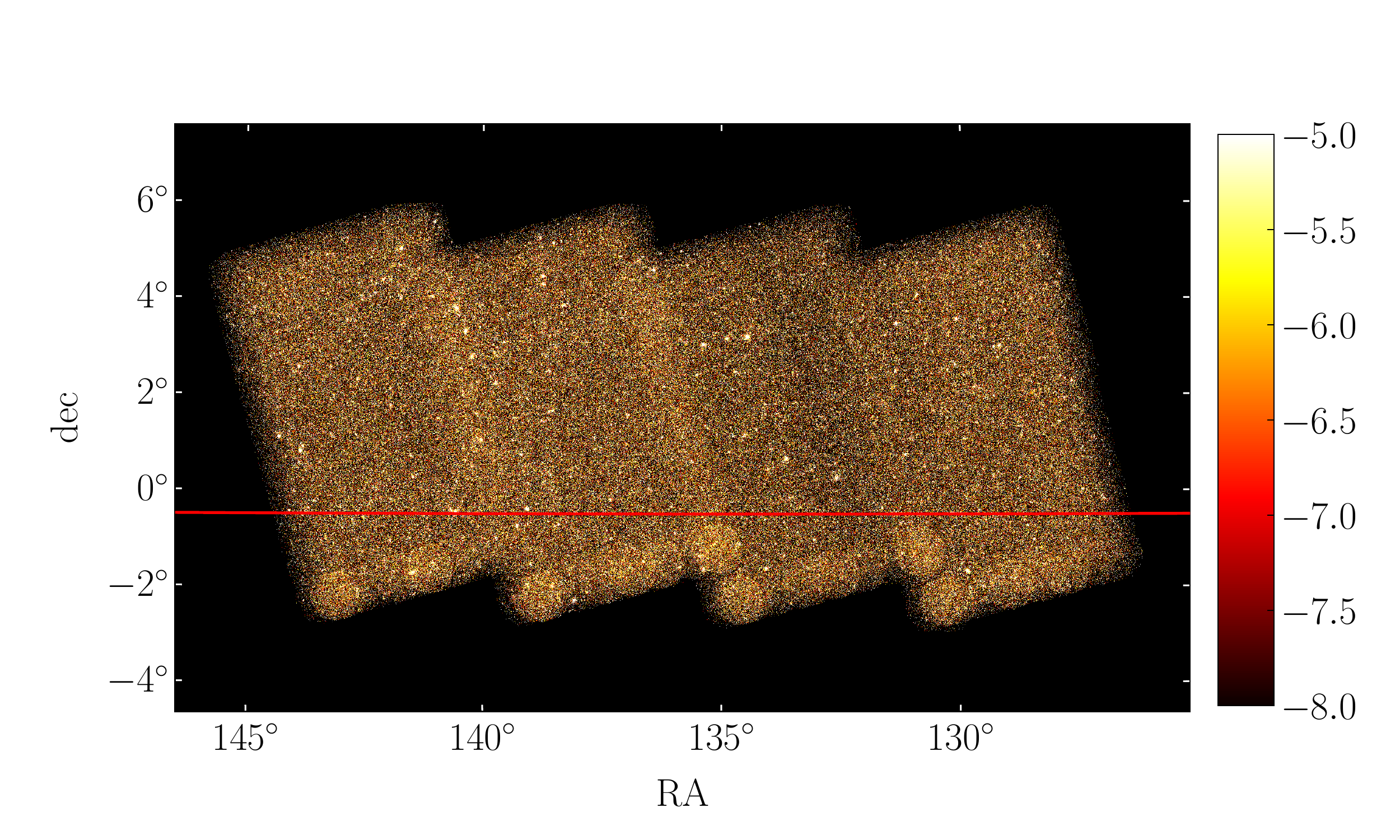}
\caption{The eFEDS photon count rate map (in counts per second) in the $[0.5, 2.0]$~keV band. The scale in the colourbar is in $\log_{10}$ scale. The red line indicates declination $\delta = -0.5\degree$ below where there are visible artefacts. We measured the power spectra with and without this region.  \label{fig:eFEDS_map}}
\end{figure*}

\subsection{Construction of the eFEDS surface brightness map}\label{sec:efeds_map}

We obtain the raw event files of the eFEDS footprint from the {\em eROSITA}-DE Early Data Release website.\footnote{\url{https://erosita.mpe.mpg.de/edr/eROSITAObservations/Catalogues/}}
The eFEDS data consists of four event files, each being a quadrangular patch with an area of 35 square degrees. We use the {eSASS} software package\footnote{\url{https://erosita.mpe.mpg.de/edr/DataAnalysis/}} to perform data reduction and analysis, following \citet{chadayammuri_etal22}. For each event file, we use the {\tt evtool} command to select all photons with energy between $0.5$ and $2.0$~keV and convert them into an image count map, with a pixel size of $4$~arcsec. We generate the corresponding exposure maps using the {\tt expmap} routine. 

Then, we use the {\tt CIAO} command {\tt reproject\_image} to reproject the four images onto the same WCS coordinates, and merge them together into a single field for both the count maps and exposure maps. We exclude detectors T5 and T7 from our analyses, because these were known to have contamination due to defective optical shields (see \citealt{chadayammuri_etal22} for details).   

We divide the counts image by the exposure map to get the count rate map. We then divide the count rate map by the solid angle of the pixels ($4\times 4\,{\rm arcsec}^2$) to get the X-ray surface brightness in ${\rm counts\, s^{-1}\, sr^{-1}}$. Fig.~\ref{fig:eFEDS_map} shows the eFEDS count rate map, which we use to measure the angular power spectrum. We convert the count rate from ${\rm counts\,s^{-1}}$ to physical energy units ${\rm erg\,cm^{-2}\,s^{-1}}$, using the energy conversion factor of $1.974 \times 10^{-12} {\rm erg\,cm^{-2}\,counts^{-1}}$, assuming an APEC model with $0.4$ solar metallicity and a gas temperature of $10^7\,{
\rm K}$. 

Resolved point sources are masked out in the X-ray surface brightness map. The point source catalogue we used is given in \cite{eFEDS_ptsrc}. Following \citet{chadayammuri_etal22}, each point source is masked at its position with a circular mask with radius given by the {\tt APE-RADIUS} parameter in the source catalogue, which is the instrumental point spread function (PSF). To ensure that contributions from all resolved point sources are masked, we use a masking radius of three times {\tt APE-RADIUS}, which is on average $\sim 0.66$ arcmin.  In addition, we make sure that central regions of the extended sources from the eFEDS extended source catalogue that were marked erroneously as point sources are unmasked.  

We then interpolate the count rate map in HEALpix pixels \citep{healpix}, with $N_{\rm side} = 8192$, with the {
\tt reproject} python package. The corresponding angular size of the HEALpix pixel is $0.25\arcmin$ (compared to PSF of $0.18\arcmin$). Since eFEDS has a fairly large coverage of 140~deg$^2$, we avoid potential biases of making the flat-sky assumption by using HEALpix projection. The HEALpix interpolated map also allows us to take advantage of the HEALpix analysis package in computing the angular power spectrum. 

\subsection{Excluding artefacts in the eFEDS field}

The eFEDS field shown in fig.~\ref{fig:eFEDS_map} shows circular features with enhanced surface brightness for declination $\delta <-0.5$\degree that are unnatural. We believe they are artefacts of the instrument or systematics due to scanning strategy. To assess these features from contaminating our power spectrum measurements, we also measure the power spectrum excluding the regions with $\delta <-0.5$\degree.

\subsection{Measuring the Angular Power Spectrum}\label{sec:measurments}

The first step to obtain the angular power spectrum is to measure the ``pseudo'' angular power spectrum, given by
\begin{equation}
    \tilde{C}_\ell = \frac{1}{2\ell+1}\sum^{\ell}_{m=-\ell} a_{\ell m}^*a_{\ell m},
\end{equation}
where $\ell$ is the multipole, which is related to a given angular scale $\theta$ by $\ell \sim \pi/\theta$. Here $a_{\ell m}$ is the coefficient of the spherical harmonics corresponding to mode $(\ell,m)$ and $a_{\ell m}^*$ is its complex conjugate. We use the {\tt anafast} function from {\tt healpy} \citep{healpy} to compute $\tilde{C}_\ell$ from multipole $\ell=1$ to $\ell_{\rm max} = 3N_{\rm side}-1 = 24575$. 
The relationship between the pseudo angular power spectrum and $\tilde{C}_{\ell}$ and the true angular power spectrum ${C}_\ell$ is given by  
\begin{equation}
    \tilde{C}(\ell) = \sum_{\ell'} M(\ell,\ell') C(\ell')B(\ell')^2 + {N},
\end{equation}
where $B(\ell')$ is the beam transfer function that account for the effect of the PSF of the instrument, $M(\ell,\ell')$ is the mode coupling matrix, which describes the coupling of power at different multipoles due to masking and the limited sky coverage, and ${N}$ is the power spectrum due to shot noise. 

The beam transfer function is $B(\ell) = \exp(-\ell(\ell+1)\sigma^2/2)$, where $\sigma = \theta_{\rm FWHM} /\sqrt{8\ln 2}$ and $\theta_{\rm FWHM} = \, 18$~arcsec is angular PSF of the instrument. We assume the beam to be Gaussian. 

The mode coupling matrix $M(\ell,\ell')$ can be computed analytically \citep[e.g.,][]{hivon_etal02} or via Monte-Carlo simulations. We adopt the latter method which is computationally less intensive. We generated 100 mock X-ray surface brightness maps with the same footprint as eFEDS, and their corresponding full sky counterpart using an input model power spectrum based on the ICM model in Section~\ref{sec:model}. Next, we construct our mock maps using the {\tt synfast} function in the {\tt healpy} package. For each map, we measured the power spectra of the eFEDS fields and the full sky map, and we took the average of their differences over the 100 map realizations. The average power spectrum over the mock maps is smaller than the input model power spectrum by the sky fraction $f_{\rm sky} = 140\,{\rm deg}^2 / (4\pi) = 0.00339$, where the mock power spectra are smaller than the input by $f_{\rm sky}$. Once we accounted for the sky fraction, we found that there are negligible differences between the power spectra of the eFEDS field and the full-sky field at multipoles greater than 100. The errors on the power spectrum are greater than the difference between the two fields at these multipoles, indicating that mode coupling does not play a significant role at these scales. 

While resolved point sources were masked in our eFEDS surface brightness map, the population of unresolved point sources will still contribute.  Since these unresolved sources are mainly clustered below the angular resolution and thus contribute to the shot noise power spectrum $N$, which needs to be subtracted. As we expect the shot noise to dominate at the smallest angular scales, we determine the total shot noise power spectrum $N$ by measuring the mean power at the 10 highest $\ell$ bins \citep[see appendix C in][]{kolodzig_etal17}.

We then bin the multipoles in 51 logarithmically spaced bins from $\ell = 20$ to $\ell = 20000$. We estimate the $1\sigma$ variance of $C_\ell$ with
\begin{equation}\label{eq:variance}
    (\Delta C_\ell)^2 = \frac{2}{2\ell+1} \frac{\left (C_\ell+ N\right)^2}{f_{\rm sky}\Delta \ell}, 
\end{equation}
where $\Delta \ell$ is the multipole bin width. 

\subsection{Power due to unresolved point sources}

We estimate the contribution of power due to clustering of unresolved AGNs using the model presented in \citet{helgason_etal14} by assuming a step selection function with a flux limit for the eFEDS of $10^{-14}\,{\rm erg\,s^{-1}\,cm^{-2}}$.

\section{Halo Model Power Spectrum}\label{sec:model}

\subsection{Modelling the X-ray auto power spectra of clusters and groups } \label{s:M_APS}

We compare the eFEDS X-ray power spectrum with the halo model. In the halo model, the X-ray power at a given angular scale $\ell$ is given by
\begin{eqnarray}\label{eq:cl_icm}
C_{\ell} &=& C^{\mathrm{1h}}_{\ell} + C^{\mathrm{2h}}_{\ell}, \\
C^{\mathrm{1h}}_{\ell} &=& \int_0^{z_{\rm max}} dz\,
\frac{d^2 V}{dzd\Omega} \nonumber \\
& \times & \int_{M_\mathrm{min}}^{M_\mathrm{max}} dM\, \frac{dn}{dM}
|S_\ell (M, z)|^2 , \\
C^{\mathrm{2h}}_{\ell} &=& \int_0^{z_{\rm max}} dz\, \frac{d^2 V}{dzd\Omega}
P_\mathrm{m} (k_\ell , z) \nonumber \\
& \times & \left[
\int_{M_\mathrm{min}}^{M_\mathrm{max}} dM\, \frac{dn}{dM}
b(M, z) S_\ell(M, z)
\right]^2 ,
\end{eqnarray}
where $k_\ell = \ell/\{ (1+z)d_A(z) \}$, $z_{\rm max}$ is the maximum redshift of observed clusters and groups for a given flux-limited X-ray survey, $d_A(z)$ is the angular diameter distance, $d^2 V/dzd\Omega = (1+z)^2 d_A^2/H(z)$ is the comoving volume per redshift and solid angle. Note that the one-halo term $C^{\mathrm{1h}}_{\ell}$ dominates over the the two-halo term $C^{\mathrm{2h}}_{\ell}$ for the angular scales probed by eFEDS $\ell > 100$. 
We use the Tinker halo mass function $dn/dM$ and halo mass bias function $b(M,z)$ \citep{tinker_etal08,tinker_etal10} to model the abundance and spatial distributions of dark matter halos, and adopt the flat $\Lambda$CDM cosmological parameters from \citet{planck2018}, with $H_0 = 67.3\,{\rm km/s/Mpc}$, $\Omega_m = 0.315$, $\Omega_b = 0.049$, and $\sigma_8 = 0.8159$, unless noted otherwise. For the same cosmology, we used the linear matter power spectrum $P(k)$ computed from {\tt CAMB}\footnote{\url{https://camb.info}}, a cosmology code for calculating matter power spectra and transfer functions.  

We choose the lower and upper limits of the halo mass to be $(M_\mathrm{min},M_\mathrm{max}) = (10^{13}, 10^{16})\,M_\odot$, covering group and cluster mass scales, and $z_{\rm max} = 2.0$ throughout the paper. Almost all of the contribution of the X-ray angular power spectrum comes from halos with mass and redshift within these ranges \citep[see fig. 1 in][]{shirasaki_etal19}. 

The term $S_{\ell}(M,z)$ is the Fourier transform of the X-ray surface brightness profile of halo with mass $M$ and redshift $z$ (c.f. Equation~\ref{eq:xsb}):
\beq
\label{eq:S_ell}
S_\ell = \frac{4\pi R_{500}}{\ell_{500}^2}
\int dx\, x^2 s_X(x;z) \frac{\sin (\ell x/\ell_{500})}{\ell x/\ell_{500}},
\eeq
where $x = r/R_{500}$, $\ell_{500} = d_A/R_{500}$, $R_s$ is the scale radius.
The model X-ray emissivity profile $s_X$ for a given halo is computed as
\beq\label{eq:xsb}
s_X(r;z) = \frac{n_H(r) n_e(r)}{4\pi(1+z)^4} \int^{E_{\rm max}(1+z)}_{E_{\rm min}(1+z)}\Lambda(T(r),Z,E) dE , 
\eeq
where $n_H$, $n_e$ and $T$ are the hydrogen and electron number densities and gas temperature respectively. We use the {\tt APEC} plasma code version 3.0.9 \citep{foster_etal12} to compute the X-ray cooling function $\Lambda$, integrated over energy range at $z=0$: $[E_{\rm min}, E_{\rm max}] = [0.5, 2.0]$~keV in the observer's frame, same as the eFEDS data. We assume constant metallicity of $Z=0.3 Z_\odot$ throughout the ICM, as suggested from observations \citep[e.g., see][for review on ICM metallicity]{mernier_etal18}.

\subsection{Halo gas model}\label{sec:gas_model}

\begin{table*}
\begin{center}
\begin{tabular}{ |c|l|l| }
\hline
Parameter &  Physical meaning & Fiducial Value \\ 
\hline \hline
$10^6\epsilon_f$ & feedback efficiency from SNe and AGNs & $3.97$ \\ 
$\sqrt{\mathcal{C}}f_{\rm gas}$ & amplitude of gas mass fraction biased by clumping &  $0.011$ \\ 
$S_{\rm gas}$ & mass slope of gas mass fraction & $0.23$ \\
$\Gamma_{\rm mod0} $ & polytropic index within cluster core $r/R_{500c}<0.2$ & $0.1024$ \\
$A_{\rm nt}$ & amplitude of non-thermal pressure fraction profile & $0.451$ \\ 
\hline
\end{tabular}
\end{center}
\caption{Values of the ICM model parameters. They are taken from the best fits of the gas density profiles from {\em Chandra}-SPT cluster samples from \citet{flender_etal17}, except the parameters for the non-thermal pressure fraction profile and gas density clumping profile, which are taken from the fit to the {\em Omega500} cosmological simulation \citep{nelson_etal14}.
\label{tab:params}
}
\end{table*}

The profiles of $n_H$, $n_e$ and $T$ in Equation~\ref{eq:xsb} are computed with the Baryon Pasting (BP) halo gas model \citep{bp_algo}. The halo gas model is described in \citet{shaw_etal10} and \citet{flender_etal17}, which we discussed in more detail in \citet{shirasaki_etal19}. This model has been used to constrain ICM astrophysics and cosmology with tSZ lensing cross-correlations \citep{osato_etal18, osato_etal20}, and mock {\em eROSITA} simulations \citep{comparat_etal20}. Here we overview the salient features of this model. 

The model assumes that the dark matter density profile of the halo follows the Navarro-Frenk-White (NFW) profile \citep{nfw96}, which is specified completely by halo redshift, the virial mass of the halo $M_{\rm vir}$, and the halo concentration parameter $c_{\rm vir} = R_{\mathrm{vir}}/R_s$ where $R_{\mathrm{vir}}$ and $R_s$ are the virial radius and the NFW scale radius, respectively. We use the formulation in \citet{diemer_kravtsov15} to compute the halo concentration for a halo with given mass and redshift. 

The {\em total} gas pressure (thermal + non-thermal) $P_{\rm{tot}}$ is then assumed to be in hydrostatic equilibrium with the gravitational potential of the NFW DM halo in our model. The relationship between the total pressure and the gas density $\rho_{g}$ is related through a polytropic relation, with $P_{\rm{tot}}(r) = P_0 \theta(r)^{n+1}$, where
$\rho_{g}(r) = \rho_0 \theta(r)^{n}$, and 
$\theta(r) = 1 + \frac{\Gamma-1}{\Gamma}\frac{\rho_0}{P_0}(\Phi_0 - \Phi(r))$. Here $\theta$ is a dimensionless function that represents gas temperature, $\Phi_0$ is the central potential of the cluster given by the NFW profile, and ${\Gamma=1+1/n}$ is the polytropic exponent, a parameter in our model. 
Cosmological hydrodynamical simulations suggest that $\Gamma\approx 1.2$ outside cluster cores where $r> 0.2 R_{500c}$ \citep{komatsu_seljak02, ostriker_etal05,shaw_etal10,battaglia_etal12}. Recent observations show consistent values for this value of the polytropic index \citep{ghirardini_etal19}. 

The normalization constants $P_0$ and $\rho_0$ are then determined numerically by solving the energy and momentum conservation of the ICM. In particular, the energy of the ICM gas is given by
\beq \label{eq:E}
E_{g,f}  =  E_{g,i} + \epsilon_f M_\star c^2 + \Delta E_p.
\eeq
where $E_{g,f}$ and $E_{g,i}$ are the final and initial total energies (kinetic plus thermal plus potential) of the ICM. $\Delta E_p$ is the work done by the ICM as it expands.  
The term $\epsilon_f M_\star c^2$ is the energy injected into the ICM due to feedback from both supernovae (SNe) and active galactic nuclei (AGNs), where $M_\star$ is the total stellar mass. 
Since the X-ray power spectrum is sensitive to the mass of the X-ray emitting gas, rather than the stellar mass in the halo, we modify our model to use the gas mass fraction instead of the stellar mass fraction. The gas mass fraction $F_{\rm gas}$ is modelled as
\beq
F_{\rm gas} (M_{500}) = f_{\rm gas}\left(\frac{M_{500}}{3\times10^{14}M_\odot}\right)^{S_{\rm gas}}.
\eeq
which is described by two parameters $(f_{\rm gas}, S_{\rm gas})$ that control the normalization and the slope of the $F_{\rm gas}-M$ relation. The stellar mass fraction is simply given by subtracting the gas mass fraction from the cosmic baryon fraction $F_{\star}  = F_{\rm baryon} - F_{\rm gas}$. 

Since the X-ray emission from the hot cluster gas is proportional to density squared (as the emission is a two-body process between electron and ion), the X-ray derived gas density will be biased by the gas density clumping factor $\mathcal{C} = {\langle n_{\rm gas} ^2 \rangle}/{\langle n_{\rm gas}\rangle^2} \geq 1$. 
The clumping factor quantifies the level of ``clumpiness'' in gas density. Cosmological simulations have shown that $\mathcal{C}\gg 1$ in cluster outskirts \citep{nagai_lau11, vazza_etal13, zhuravleva_etal13}. Failing to account for density inhomogeneity will lead to overestimates in gas density and mass \citep{mathiesen_etal99} by the square root of the clumping factor $\sqrt{\mathcal{C}}$, and also underestimates in hydrostatic mass as it is inversely proportional to gas density \citep{roncarelli_etal13}. 

Clumping can affect X-ray power spectrum through its gas mass fraction $F_{\rm gas}$ dependence. Since with power spectrum alone we will not be able to separately measure the clumping factor, the gas mass fraction we constrain is affected by gas clumping. We therefore merge the clumping dependence into the gas mass fraction parameters $\sqrt{\mathcal{C}}f_{\rm gas}$. Here $\mathcal{C}$ represents the gas clumping factor computed within $R_{500c}$. 
Recent observations of nearby clusters place an upper limit on the clumping factor at $\sqrt{\mathcal{C}} \lesssim 1.4$ out to $R_{200c}$ \citep[e.g.][]{eckert_etal15,tchernin_etal16,morandi_etal17, mirakhor_walker21}. 

The model also assumes that the ICM in the dense core of the galaxy cluster follows a different polytropic equation of state than the rest of the ICM due to strong cooling and feedback in the core. Following \citet{flender_etal17}, we use three parameters to describe the physical state of the ICM in cluster cores: $\xbr=r_{\rm break}/R_{500}$ where the adiabatic index changes, a separate adiabatic index $\Gamma_0$ for $x<\xbr$, and ${\beta_g} = d\ln \Gamma_0 /d\ln (1+z)$  models the redshift dependence of $\Gamma_0$. 

We include the effects of non-thermal pressure in the ICM by adopting the ``universal'' non-thermal pressure fraction profile from \citet{nelson_etal14b}:
\begin{equation}
\frac{P_{\rm nt}(r)}{P_{\rm th}(r)+P_{\rm nt(r)}} =  1 - A_{\rm nt}\left[1+\exp\left\{-\left(\frac{r}{B_{\rm nt}\, R_{ 200m}}\right)^{C_{\rm nt}}\right\}\right]. \label{eq:nelson}
\end{equation}
The model is specified by three parameters $A_{\rm nt}$, $B_{\rm nt}$, and $\gamma_{\rm nt}$, which represent the normalization, radial dependence, and shape of the non-thermal pressure fraction profile, respectively. These parameters are not well constrained observationally \citep[see][]{eckert_etal18}, thus we take the best-fitting values from the {\em Omega500} cosmological simulations \citep{nelson_etal14} with $A_{\rm nt} = 0.451, B_{\rm nt} = 0.841$, and $\gamma_{\rm nt} = 1.628$. Note that the non-thermal pressure only changes the relative contribution of non-thermal vs thermal pressure in the ICM, thus the X-ray angular power spectrum, which mainly depends on density with weak temperature dependence, is only weakly sensitive to the non-thermal pressure fraction. 

The model parameters are summarized in Table~\ref{tab:params}.

\subsection{Dependence of the X-ray Angular Power Spectrum on Model Parameters}\label{sec:model_dependence}

\begin{figure*}
\centering
\includegraphics[width=2.\columnwidth]{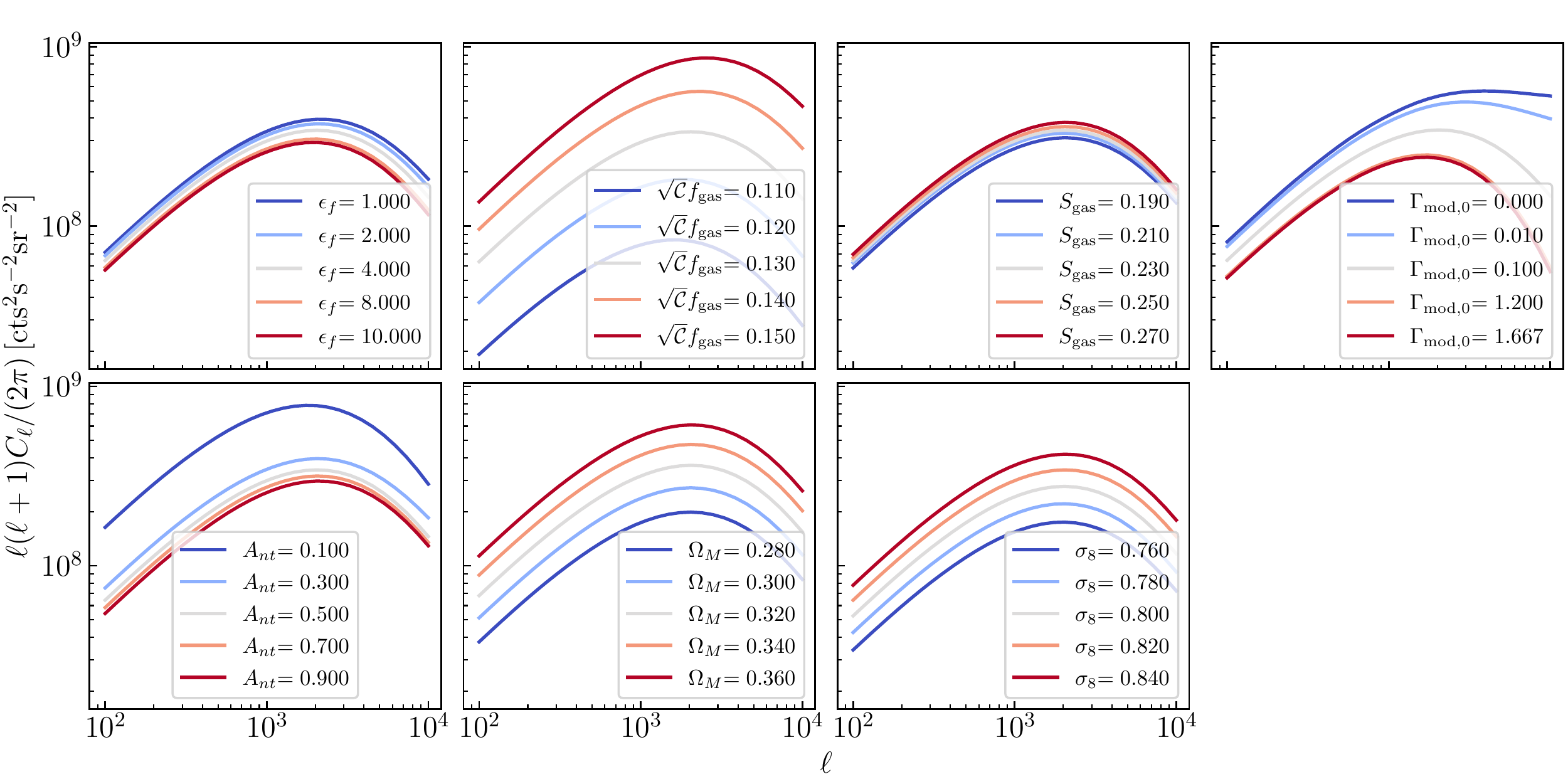}
\caption{Plots showing how the X-ray angular power spectrum depends on the selected parameters of the ICM model listed in Table~\ref{tab:params}, as well as $\Omega_M$ and $\sigma_8$. In each panel, only one parameter is varied, while the other are fixed to the fiducial values . \label{fig:param_xx_power}
}
\end{figure*}

In fig.~\ref{fig:param_xx_power}, we show the dependence of the model X-ray angular power spectrum on the key model parameters: the feedback efficiency $\epsilon_f$, the normalization $f_\star$ and slope $S_\star$ of the stellar mass fraction, the polytropic exponent in the inner halo core $\Gamma_{\rm mod,0}$, the amplitudes of the non-thermal pressure fraction $A_{nt}$, as well as cosmological pararmeters $\Omega_M$ and $\sigma_8$.   

Increasing the feedback efficiency $\epsilon_f$ decreases gas density inside halo cores, leading to lower X-ray surface brightness (which depends on gas density squared). 

Similarly, increasing the gas mass fraction normalization $\sqrt{\mathcal{C}}f_{\rm gas}$ increases the overall amount of gas mass inside halos, which also leads to higher X-ray emission. Increasing $S_{\rm gas}$ increases the amount of gas mass in more massive halos with larger angular sizes, thus increases the power spectrum normalization at all scales. 

Lowering the polytropic index in the halo core $\Gamma_{\rm mod,0}$ results in denser halo core gas, which increases X-ray power at small angular scales. 

Altering the amplitude of the non-thermal pressure fraction $A_{nt}$ also changes the X-ray power spectrum, by changing the temperature of gas in halo. Halos with high non-thermal pressure do not have gas with temperature high enough to be X-ray emitting. 
Increasing $\Omega_M$ increases the amount of gas in halos at fixed cosmic baryon fraction, while higher $\sigma_8$ increases the number of clusters and groups at a given mass and redshift, both leading to increases in the normalization of the X-ray angular power spectrum. 

\section{Results}
\label{sec:results}

\subsection{The eFEDS power spectrum}

\begin{figure*}
\centering
\includegraphics[width=2\columnwidth]{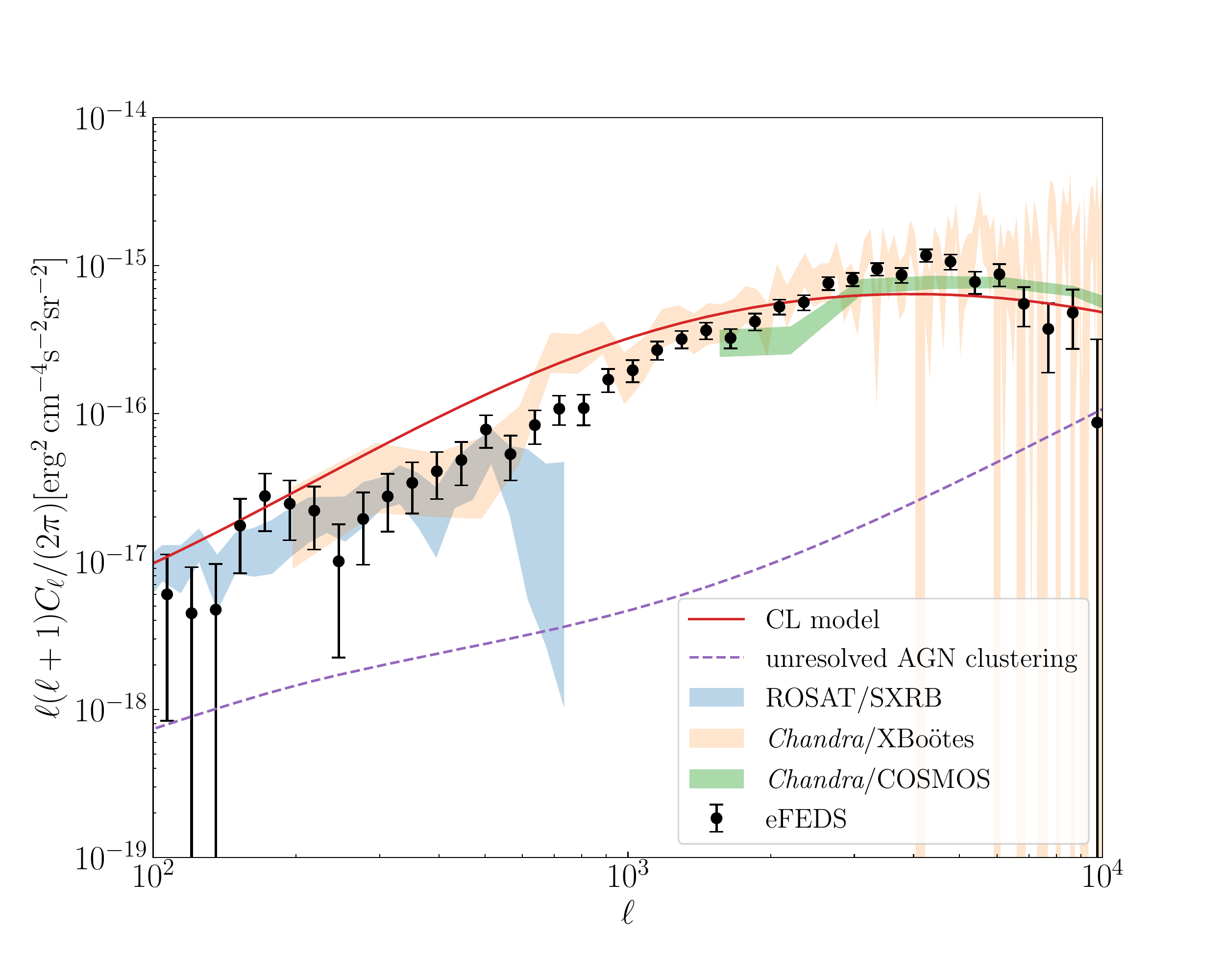}
\caption{The black data points represent the X-ray angular power spectra of extended sources measured in the eFEDS field with $1\sigma$ errorbars. We also show the $\pm
1\sigma$ range of the power spectra from three other measurements: {\em ROSAT}/SXRB in blue, {\em Chandra}/XBo\"otes in orange, and {\em Chandra}/COSMOS in green. The dashed line shows the model power spectrum from clustering of unresolved AGNs, and the solid line shows our halo model prediction.} 
\label{fig:eFEDS_power} 
\end{figure*}

Fig.~\ref{fig:eFEDS_power} shows the resulting X-ray angular power spectrum measured from the eFEDS field with the resolved point sources masked, with both shot noise and unresolved AGN power subtracted. The power spectrum contribution of clustering of unresolved AGNs is subdominant at all angular scales. Excluding the region with instrumental artefacts ($\delta < -0.5\degree$) does not change the measured power spectrum significantly. The artefacts lead to structures at large angular scales ($\ell \lesssim 100$) that are dominated by cosmic variances.  

\subsection{Comparison of eFEDS with halo model power spectrum}\label{sec:model_compare}

We compare the eFEDS power spectrum with our halo model prediction. The model in general agrees with eFEDS measurements at angular scales smaller than $\ell \sim 2000$, corresponding to the scale where shot-noise starts to dominate. At larger $\ell$, the eFEDS power is larger than the model prediction. The excess is unlikely caused by clustering of unresolved point sources, as the predicted value is small. We believe that the excess could be due to incomplete masking of resolved point sources as we discuss in Section~\ref{sec:cosmos}. 

\subsection{Comparison with the power spectrum measurements of {\em ROSAT}/SXRB}\label{sec:ROSAT}

We compare the power spectrum of eFEDS with that of the Soft X-Ray Background (SXRB) from the {\em ROSAT} All Sky Survey \citep{snowden_etal97}. We used the {\em ROSAT}/SXRB in HEALpix format (with $N_{\rm side} = 512$) for analysis\footnote{The maps are available at  \url{http://www.jb.man.ac.uk/research/cosmos/ROSAT/}}. We use the combined R4 and R7 maps for the analysis, which covers the energy range of $[0.4,2.0]$~keV. We excluded most of the Galactic emissions by masking out regions with Galactic latitude between $l \in [-40\degree, 40\degree]$, and Galactic longitude between $b \in [80\degree, 250\degree]$, with sky fraction $f_{\rm sky} = 0.18$. We follow the same power spectrum estimation procedure described in Section~\ref{sec:measurments} for the {\em ROSAT}/SXRB measurements. The {\em ROSAT}/SXRB power spectrum is limited to $\ell < 1000$ because of the poor angular resolution of 12 arcmin in SXRB. We find excellent agreement between {\em ROSAT}/SXRB and eFEDS at these scales. Note that while the errorbars of the {\em ROSAT}/SXRB power spectrum at $\ell < 200$ are smaller than that of eFEDS, as {\em ROSAT}/SXRB has a larger $f_{\rm sky}=0.18$, compared to $0.034$ of eFEDS, the {\em ROSAT}/SXRB errorbars are dominated by poisson noise at larger $\ell$. It is expected that the full sky eRASS will outperform {\em ROSAT}/SXRB with smaller errorbars due to lower poisson noise. 

\subsection{Comparison with the power spectrum measurements of {\em Chandra}/COSMOS}\label{sec:cosmos}

We also compare the eFEDS power spectrum with the that of the 2.2~deg$^2$ field of the {\em Chandra} COSMOS Legacy Survey \citep{cosmos}. With the high angular resolution of {\em Chandra} and with relatively deep exposure with an average of 160 ks exposure across the field, the COSMOS field has an extensive catalogue of point sources that can be used to evaluate the contribution of unresolved point sources in eFEDS to the power spectrum of extended sources. 
Following the same power spectrum measurement procedure as in \citet{li_etal18}, we measure the power spectrum for both resolved and unresolved extended sources in the COSMOS field. Given the limited field of view, we are only able to reliably measure the angular power for multipoles $\ell > 3000 $. 
The eFEDS data shows consistent power spectrum with the COSMOS data. The consistency means that although eFEDS data has lower angular resolution, point sources are effectively masked such that it reproduces similar power as the high angular resolution {\em Chandra}/COSMOS measurements. 

\subsection{Comparison with the power spectrum measurements of {\em Chandra}/XBootes}\label{xbootes}

We compare the eFEDS measurements with the X-ray angular power measurements in the {\em Chandra} XBootes field \citep{XBootesI}. The {\em Chandra}/XBootes field has an area of 9 deg$^2$ with an average exposure time of 5~ks. We refer the reader to \citet{kolodzig_etal17,kolodzig_etal18} for details on the power spectrum measurements. The eFEDS power spectrum shows reasonable agreement with that of {\em Chandra} at $\ell \in [700,3000]$.  At $\ell \lesssim 700$, corresponding to angular scales of $\theta > 0.25$ degrees, the power in the {\em Chandra}/XBootes spectrum underestimates slightly the clustering power, as it suffers from sample variance due to its relatively small footprint. At higher $\ell > 3000$, the error on the {\em Chandra}/XBootes power becomes too large to make meaningful comparison with eFEDS.  

\subsection{Cosmological forecasts for the {\em eROSITA} All Sky Survey}

\begin{figure*}
\centering
\includegraphics[width=1.9\columnwidth]{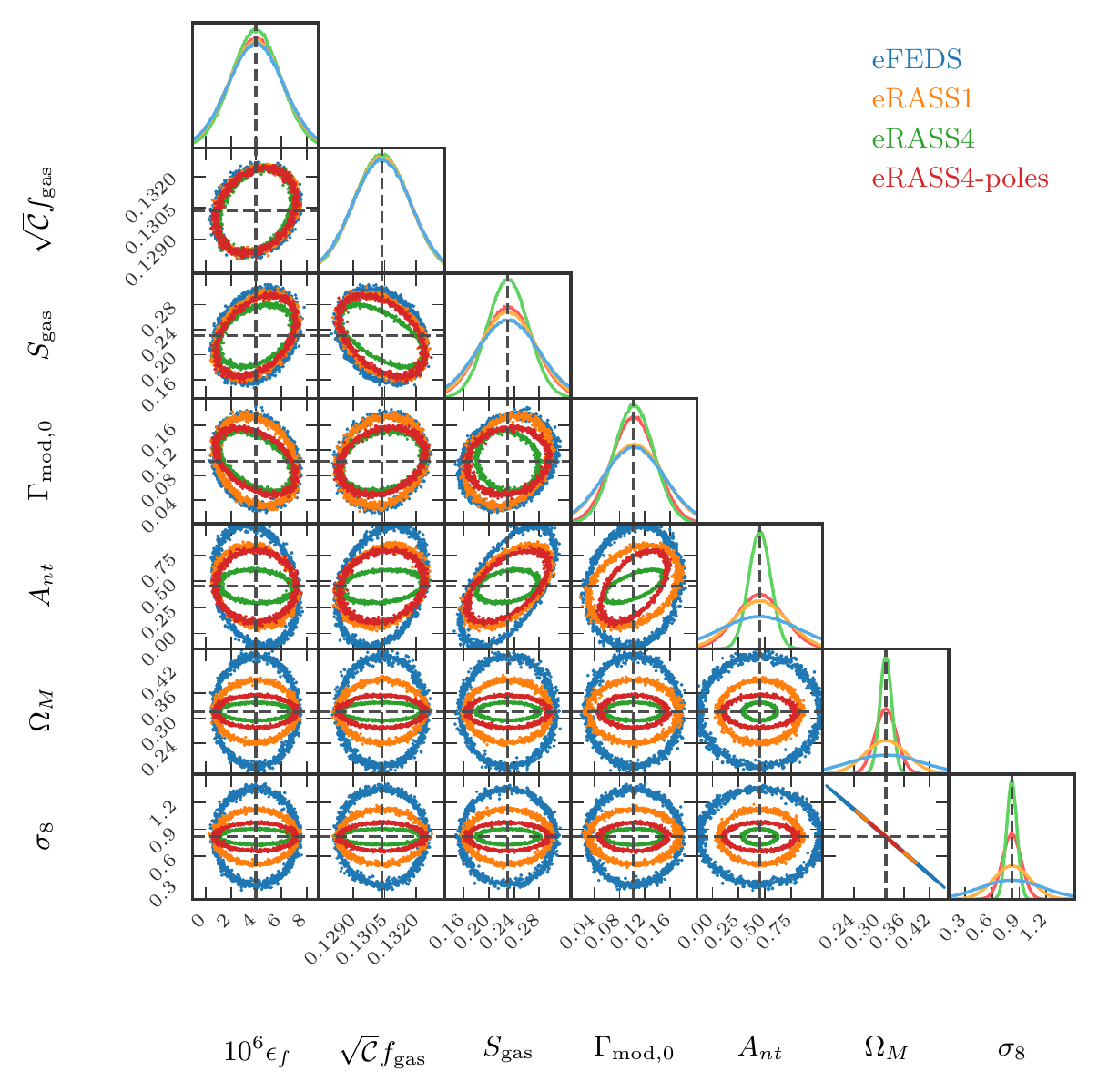}
\caption{Plot showing the Fisher forecasts of the ICM parameters and $\Omega_M$ and $\sigma_8$ for the eFEDS, eRASS1, eRASS4, and eRASS4-poles survey. The ellipses denote $68\%$ confidence contours of the parameters. 
\label{fig:fisher} }
\end{figure*}

We perform a Fisher forecast  to predict how future all-sky X-ray surveys are able to constrain cluster astrophysics and cosmology. The surveys we consider here are the first (eRASS1) and fourth (eRASS4) data release of the full-sky scan\footnote{At this point, it is not clear whether the final scan, eRASS8, will be completed. }. The eRASS1 survey will have approximately a minimum exposure time of 200 ks across the whole sky, about 1/8th of that of eFEDS,  while eRASS4 will have and exposure time of 800 ks, about half of that of eFEDS. We consider a sky coverage of $40\%$ for both eRASS1 and eRASS4, for half of the sky that will be made public (from the German team), and taking into account the masking of the Milky Way Galaxy and other X-ray foregrounds. 
We compute the Fisher matrix for parameters ${p_i, p_j}$ as
\beq
F_{ij} = \sum_{\ell,\ell'} \frac{\partial C_{\ell}}{\partial p_i}\Sigma^{-1}(\ell, \ell')\frac{\partial C_{\ell'}}{\partial p_j} \delta_{\ell, \ell'},
\eeq
where $\Sigma$ is the covariance matrix in the power spectrum measurements we use the same expression as in Equation~\ref{eq:variance}. For simplicity, we assume that off-diagonal elements are negligible e.g., no mode coupling or mixing of scales in the measurements. This is justified by the low level of correlated errors in the power spectrum measurements at the interested multipole scale $\ell \in [100, 10000]$ in 50 logarithmic bins. 
We also add $1\sigma$ Gaussian priors for the gas physics parameters $\{ \epsilon_f, \sqrt{\mathcal{C}}f_{\rm gas}, S_{\rm gas}\} $, with $\sigma = \{4.82, 0.003, 0.11\}$, respectively, from the constraints from the SPT density profile analysis from \citet{flender_etal17}. 

The minimal covariance matrix for given two parameters is then the inverse of the Fisher matrix:
\beq
Cov(p_i, p_j) = F^{-1}_{ij}. 
\eeq

Fig.~\ref{fig:fisher} shows our Fisher Forecasts for the astrophysical and cosmological parameters for eFEDS, eRASS1 and eRASS4. 
First eFEDS will not be able to provide any meaningful cosmological constraints due to its small sky coverage. This is consistent with our attempt in fitting the eFEDS power spectrum with our model, which yielded poor constraints on both the astrophysical parameters and $\Omega_M$ and $\sigma_8$.   
While eRASS1 have larger sky coverage than eFEDS, its constraining power does not improve much beyond eFEDS because of its low exposure time, leading to higher shot noise. Finally, the currently projected final release of the eRASS data, eRASS4, will be able achieve improved cosmological constraints, giving marginalized constraints on $\sigma_8$ and $\Omega_M$ at $\sim 10\%$. The ecliptic polar regions in eRASS4 (   ``eRASS4-poles'') will long exposure time of 10ks, compared to 1.6ks of eFEDS, but with limited sky coverage of $\sim 100$ square degrees for each polar regions. We find that eRASS4-poles will not provide competitive constraints as that of the half-sky eRASS4.  

We predict marginalized constraints on $\Omega_M$ and $\sigma_8$, with $2\sigma$ constraints at $1\%$ and $2\%$ respectively. At the same time, we can achieve $2\sigma$ constraints on the normalization of the gas mass fraction at $36\%$, and the normalization of non-thermal pressure fraction at $50\%$.  

Note that these constraints will be improved compared to those obtained in existing cluster counts in X-ray, Microwave, and Optical cluster surveys \citep{vikhlinin_etal09, mantz_etal10, benson_etal13, mantz_etal22}, the constraints from cross X-ray and SZ power spectrum measurements with ROSAT-{\em Planck} \citep{hurier_etal15}; and they will be comparable with upcoming cluster count constraints from eRASS \citep{pillepich_etal18}.

\section{Summary and Discussion}
\label{sec:summary}

We present the first measurement of the X-ray angular power spectrum of extended sources in the eROSITA Final Equatorial Survey (eFEDS), which covers 140~deg$^2$ of the sky. We compare the eFEDS power spectrum measurements to model predictions. Our main findings are the following:
\begin{itemize}
    \item The power spectrum of extended sources in the eFEDS field is overall consistent with other independent power spectrum measurements at multipoles $100 <\ell <10000$. In particular, it agrees well with the power spectrum of the Soft X-ray Background map from ROSAT All Sky Survey, as well as the smaller {\em Chandra}/XBootes and COSMOS fields. 
    \item The model X-ray power spectrum, based on calibration from a {\em Chandra} measurements of density profiles of the South-Pole Telescope cluster sample, provides a reasonably good match to the eFEDS power spectrum at multipoles $ 100 < \ell < 10000$ without any fitting or fine-tuning of the parameters. However, the eFEDS power spectrum errorbars are still too large to provide meaningful constraints on $\Omega_M$ and $\sigma_8$. 
    \item Using Fisher forecast, we predict the 4th data release of the upcoming eROSITA All-Sky Survey (eRASS4) will  reduce errors on the power spectrum and provide good constraints on cluster gas mass fraction and cosmological parameters, specifically on $\Omega_m$ and $\sigma_8$, at the 10 percent level. 
\end{itemize}

X-ray power spectrum is a promising alternative cosmological probe, as it is subjected to different systematic uncertainties from other cosmological probes, such as cluster abundance measurements. It can offer complementary and independent constraints on cluster astrophysics and cosmology. 

\section*{Acknowledgements}
We thank the referee for providing useful feedback that improves the paper. We thank Masato Shirsaki for providing his original power spectrum calculation code from which this work is based on, Johan Comparat for his comments and mock eFEDS maps for testing and verification of the analysis pipeline used in the paper, Alex Kolodzig for providing his measurements of the power spectrum of extended sources in the {\em Chandra}/XBootes field, and K\'ari Helgason for providing the code to estimate clustering power from unresolved point sources. 
EL and NC were supported by the College of Arts Science of the University of Miami. \'AB and RPK, acknowledge support from the Smithsonian Institution and the Chandra High Resolution Camera Project through NASA contract NAS8-03060.  This work was supported in part by the facilities and staff of the Yale Center for Research Computing. 
This work is based on data from eROSITA, the soft X-ray instrument aboard SRG, a joint Russian-German science mission supported by the Russian Space Agency (Roskosmos), in the interests of the Russian Academy of Sciences represented by its Space Research Institute (IKI), and the Deutsches Zentrum für Luft- und Raumfahrt (DLR). The SRG spacecraft was built by Lavochkin Association (NPOL) and its subcontractors, and is operated by NPOL with support from the Max Planck Institute for Extraterrestrial Physics (MPE). The development and construction of the eROSITA X-ray instrument was led by MPE, with contributions from the Dr.\ Karl Remeis Observatory Bamberg \& ECAP (FAU Erlangen-Nuernberg), the University of Hamburg Observatory, the Leibniz Institute for Astrophysics Potsdam (AIP), and the Institute for Astronomy and Astrophysics of the University of T\"ubingen, with the support of DLR and the Max Planck Society. The Argelander Institute for Astronomy of the University of Bonn and the Ludwig Maximilians Universit\"at Munich also participated in the science preparation for eROSITA. This research made use of 
data from ROSAT, {\em Chandra}, {\em eROSITA} and 
several software packages, including astropy \citep{astropy1, astropy2}, CIAO \citep{ciao}, HEALpix \citep{healpix}, and healpy \citep{healpy}.

\section*{Data Availability}
The eFEDS data used in this paper are publicly avaiable at \url{https://erosita.mpe.mpg.de/edr/eROSITAObservations/Catalogues/}. The ROSAT data used in this paper is available at \url{http://www.jb.man.ac.uk/research/cosmos/ROSAT/}. The {\em Chandra}/COSMO data and {\em Chandra}/XBootes power spectrum data are available upon request. The scripts used to analyze the data are also available upon request.



\bibliographystyle{mnras}
\bibliography{ms} 





\bsp	
\label{lastpage}
\end{document}